\documentclass[preprint]{aastex}

\def\uK{\mu {\rm K}}
\def\deg{^{\circ}}
\def\min{^{\prime}}
\def\fdg{\hbox{$.\!\!^\circ$}}

\def\farcm{\hbox{$.\mkern-4mu^\prime$}}

\def\tem{\Delta T / T}
\def\tot{{{\delta T} \over {T}}}
\def\rms{\langle (\tem)^2 \rangle^{1/2}}
\def\um{\mu {\rm m}}
\def\uK{\mu {\rm K}}
\def\om{\Omega}
\def\omz{\Omega_0}

\def\be{\begin{equation}}\def\bea{\begin{eqnarray}}\def\beaa{\begin{eqnarray*}}
\def\ee{\end{equation}}  \def\eea{\end{eqnarray}}  \def\eeaa{\end{eqnarray*}}

\tighten

\slugcomment{KSUPT-01/5 \hspace{0.5truecm} June 2001}

\begin{document}

\title{Effects of Foreground Contamination on the Cosmic Microwave 
       Background Anisotropy Measured by MAP}
 
\author{Chan-Gyung Park\altaffilmark{1}, 
        Changbom Park\altaffilmark{1}, and
        Bharat Ratra\altaffilmark{2}}
\altaffiltext{1}{Astronomy Program, School of Earth and Environmental Sciences,
                 Seoul National University, 151-742 Korea; 
                 parkc@astro.snu.ac.kr, cbp@astro.snu.ac.kr.}
\altaffiltext{2}{Department of Physics, Kansas State University, Manhattan, 
                 KS 66506; ratra@phys.ksu.edu.}

\begin{abstract}
We study the effects of diffuse Galactic, far-infrared extragalactic source, 
and radio point source emission on the cosmic microwave background (CMB) 
anisotropy data anticipated from the {\it MAP} experiment. We focus on the 
correlation function and genus statistics measured from mock {\it MAP} 
foreground-contaminated CMB anisotropy maps generated in a spatially-flat 
cosmological constant dominated cosmological model. 

Analyses of the simulated {\it MAP} data at 90 GHz ($0\fdg3$ FWHM resolution 
smoothed) show that foreground effects on the correlation function are small 
compared with cosmic variance. However, the Galactic emission, even just from 
the region with $|b|>20\deg$, significantly affects the topology of CMB 
anisotropy, causing a negative genus shift non-Gaussianity signal. 
Given the expected level of cosmic variance, this effect can be effectively 
reduced by subtracting existing Galactic foreground emission models from the 
observed data. {\it IRAS} and DIRBE far-infrared extragalactic sources have 
little effect on the CMB anisotropy. Radio point sources raise the amplitude 
of the correlation function considerably on scales below $0\fdg5$, and 
also result in an asymmetry of the genus curve (at 2 $\sigma$)
on $0\fdg5$ scales. While the removal of bright sources above 5 $\sigma$
detection limit effectively reduces the excess correlation, 
the genus asymmetry remains significant. Accurate radio point source data 
is essential for an unambiguous detection of CMB anisotropy non-Gaussianity 
on these scales. 

When all the foreground sources are subtracted from the observed CMB anisotropy
map, non-Gaussianity of cosmological origin can be detected at the 2 $\sigma$ 
level if the measured genus shift parameter $|\Delta\nu| \gtrsim 0.02$ ($0.04$)
or if the measured genus asymmetry parameter $|\Delta g| \gtrsim 0.03$ ($0.08$)
on a $0\fdg3$ ($1\fdg0$) FWHM scale.
\end{abstract}


\section{Introduction}

Following the {\it COBE}-DMR experiment detection of anisotropy in the cosmic 
microwave background (CMB) on large angular scales (Bennett et al. 1996; 
G\'orski et al. 1996, 1998), there have been many measurements of CMB anisotropy
on angular scales down to $\sim 10\min$ (see, e.g., Lee et al. 2001; 
Netterfield et al. 2001; Halverson et al. 2001 for recent measurements 
of CMB anisotropy). These observations have begun to test cosmological models 
and provide interesting constraints on cosmological parameters (see, e.g., 
Kamionkowski \& Kosowsky 1999; Rocha 1999; Gawiser \& Silk 2000 for reviews,
and, e.g., Ratra et al. 1999a, 1999b; Rocha et al. 1999; Knox \& Page 2000; 
Wang \& Mathews 2000; Douspis et al. 2001; Podariu et al. 2001; Xu, Tegmark, 
\& de Oliveira-Costa 2001; Netterfield et al. 2001; Pryke et al. 2001; 
Stompor et al. 2001 for CMB anisotropy tests and constraints).

A satellite mission, the {\it Microwave Anisotropy Probe} ({\it 
MAP})\footnote{http://map.gsfc.nasa.gov/}, will probe the CMB anisotropy at 
frequencies of 22, 30, 40, 60, and 90 GHz with expected angular resolutions of 
$0\fdg93$, $0\fdg68$, $0\fdg47$, $0\fdg35$, and $0\fdg21$ FWHM, respectively. 
At 90 GHz the sensitivity of {\it MAP} is expected to be $\sim 35$ $\mu\rm K$ 
per $0\fdg3 \times 0\fdg3$ pixel. The high quality CMB anisotropy data 
anticipated from {\it MAP} will significantly improve constraints on many 
cosmological parameters. 

Non-CMB foreground contamination must be understood and accounted for to 
fully utilize the cosmological discriminative power of {\it MAP} CMB anisotropy 
data. Recent rapid progress in the CMB anisotropy field has stimulated much 
work on modeling foregrounds and on developing methods for removing them from
acquired data (see, e.g., Bennett et al. 1992, 1994; Brandt et al. 1994; 
Dodelson \& Stebbins 1994; Tegmark \& Efstathiou 1996; Bouchet \& Gispert 1999; Tegmark et al. 2000). There are two major potential sources for foreground 
contamination: diffuse Galactic emission and unresolved point sources.

Kogut et al. (1996a, 1996b, hereafter K96) detect cross-correlations between 
the {\it COBE}-DMR and DIRBE maps, which they ascribe to Galactic dust and 
free-free emission. Using the Saskatoon data ($\sim$ 30 to 40 GHz, $\sim 1\deg$ 
resolution, Netterfield et al. 1997) and the 19 GHz survey data ($\sim 3\deg$ 
resolution, Boughn et al. 1992), de Oliveira-Costa et al. (1997, 
1998)\footnote{
  Also see Simonetti, Dennison, \& Topasna (1996) and Gaustad, McCullough,
  \& Van Buren (1996).} 
obtain a result similar to that of Kogut et al. Hamilton \& Ganga (2001)
find similar correlations using the UCSB South Pole 1994 data ($\sim$ 30 to 40 
GHz, $\sim 1\deg$ resolution, Gundersen et al. 1995; Ganga et al. 1997). At 
higher frequencies near 100 GHz and on similar angular scales, MAX data
(Lim et al. 1996; Ganga et al. 1998) shows a similar correlation. On smaller
angular scales Leitch et al. (2000) reach a similar conclusion using the OVRO 
data ($\sim$ 30 GHz). However, no such correlation is found in the Python V 
data ($\sim$ 40 GHz, $\sim 1\deg$ resolution, Coble et al. 1999). Mukherjee et 
al. (2001) and de Oliveira-Costa et al. (2000) discuss foreground contamination 
of the lower frequency, larger scale Tenerife data. At frequencies higher than 
60 GHz there is no significant correlation between CMB maps and various 
synchrotron template maps such as the 408 MHz (Haslam et al. 1981) and the 1420 
MHz (Reich \& Reich 1988) radio survey maps (Kogut et al. 1996a; K96; de 
Oliveira-Costa et al. 1997, 1998).

CMB anisotropy experiments with subdegree resolution may be sensitive 
to radio point sources. Gawiser \& Smoot (1997) study the effect of 
far-infrared (FIR) point source contamination of CMB anisotropy data by 
using the {\it IRAS} 1.2 Jy flux-limited galaxy catalog at various angular 
resolutions and observation frequencies. Gawiser, Jaffe, \& Silk (1998a) 
derive upper and lower limits on microwave anisotropy from point sources over 
the range of frequencies 10 to 1000 GHz from recent sub-arcminute resolution 
point source observations. Their upper limit is $\rms \sim 10^{-5}$ for a 
$10\min$ beam at 100 GHz. Toffolatti et al. (1998) use a luminosity evolution 
model to make predictions for the number density distribution of extragalactic 
sources and their contribution to temperature fluctuations in each channel 
of the {\it Planck Surveyor} mission.
Sokasian, Gawiser, \& Smoot (1998) estimate the effect of bright radio point 
source contamination on {\it MAP} and {\it Planck} observations. They find 
that removing bright sources above the source detection limit (5 $\sigma 
\simeq 1$ Jy, where $\sigma$ is the rms level of CMB anisotropy measured)
reduces the contamination from $\rms \sim 10^{-5}$ to $10^{-6}$, consistent 
with the conclusion of Toffolatti et al. (1998).
Recently, the Wavelength-Oriented Microwave Background Analysis Team 
(WOMBAT, Gawiser et al. 1998b; Jaffe et al. 1999) have presented many different
all-sky high resolution CMB anisotropy contaminant foreground template maps, 
including those for Galactic dust, synchrotron, FIR point source, radio point 
source, and Sunyaev-Zel'dovich cluster emission.

In this paper we generate mock all-sky foreground-contaminated CMB anisotropy
maps for the 90 GHz channel of {\it MAP} in a spatially-flat cosmological
constant ($\Lambda$) dominated cold dark matter (CDM) cosmogony. We use these
maps to investigate how foreground contamination affects the CMB anisotropy
angular correlation function $C(\theta)$ and the topology of CMB anisotropy.
In $\S$2 we summarize various foreground contamination sources. In $\S$3 we 
summarize how mock foreground-contaminated CMB anisotropy maps are generated. 
In $\S$4 we study the effects of foreground contamination on the amplitude 
and shape of the correlation function, focussing particularly on the location 
of the acoustic valley in the function $\theta^{1/2} C(\theta)$.
In $\S$5 we use the genus statistic to study the effect of foreground
contamination on detecting non-Gaussianity of cosmological origin in the 
$MAP$ data. We conclude in $\S$6.

\section{Sources of Foreground Contamination}

\subsection{Galactic Dust and Free-Free Emissions}

In the microwave sky three Galactic emission components, those due to 
synchrotron, free-free, and dust, are important. Although these Galactic 
foregrounds can be modeled theoretically pixel by pixel by multifrequency 
observations (Brandt et al. 1994), this requires very low instrument noise 
and accurate emission models. According to Bennett et al. (1992), synchrotron 
and free-free emissions have well-determined spectral behaviors with 
$I_\nu \propto \nu^{-0.7 \sim -1.1}$ and $I_\nu \propto \nu^{-0.15}$, 
respectively. It is very difficult to produce an accurate all-sky dust emission 
model because its spectral behavior depends on the shape, composition, size 
distribution, and dynamics of the dust grains, and because the dust emission 
is a sum over the emission from all dust grains along each line of sight.

Kogut et al. (1996a) and K96 apply a constant temperature warm dust and 
free-free emission model with spectral intensity
$$   
     I_\nu = \tau ( \nu/\nu_0 )^{\beta} B_\nu (T) + A_{ff}
             ( \nu/\nu_0 )^{-0.15}    
     \eqno (1) 
$$
to the DIRBE maps and the DMR two- and four-year data. Here $B_\nu (T)$ is 
the Planck black body function, $\tau$ opacity, $\beta$ emissivity power-law
index, $\nu_0 = 900$ GHz, and $A_{ff}$ the normalized amplitude of free-free 
emission.
For the four-year DMR data they obtain $\tau = (1.2_{-0.4}^{+0.7})\times 
10^{-5}$, $T=20.0_{-4.0}^{+2.5}$ K, and $\beta=1.5_{-0.3}^{+1.1}$ (68\%
confidence limits) at $|b|>20\deg$. In eq. (1) the second term represents 
free-free emission, where the constant $A_{ff}$ is determined from the rms 
level of DMR-DIRBE correlated free-free emission, $\Delta T_{ff}=7.1 \pm 1.7$ 
$\mu$K at 53 GHz (see $\S$3 of K96).
Since synchrotron emission is dominant only at frequencies lower than
about 60 GHz, we neglect synchrotron emission in our analyses here.
Figure 1 shows the spectra of the dust plus free-free emission model of K96.
The thick solid curve represents the best-fit model. The upper and lower thin 
solid curves are the upper ($T=22.5$ K, $\beta=1.2$) and lower ($T=16.0$ K, 
$\beta=2.6$) limits of the K96 model. $\tau = 1.2 \times 10^{-5}$
in all three cases. The dust temperature is almost inversely proportional to 
the emissivity (Kogut et al. 1996a). 

However, dust emission models with a constant dust temperature everywhere on 
the sky are a poor description of reality (e.g., Finkbeiner, Davis, \& Schlegel 
1999, hereafter FDS99). It is known from FIRAS observations that there seem to 
exist both warm ($16$ K $< T < 21$ K) and cold ($4$ K $< T < 7$ K) dust 
components that contribute to Galactic dust emission (Wright et al. 1991; 
Reach et al. 1995). FDS99 provide a two-component temperature-varying dust 
model determined by fitting 100 $\mu$m dust emission data (Schlegel, 
Finkbeiner, \& Davis 1998, described below) to FIRAS dust spectra. 
Their model parameter values are $\alpha_1=1.67$, $\alpha_2=2.70$, 
$f_1=0.0363$, and $q_1/q_2=13.0$, where $\alpha_i$ is the $i^{\rm th}$ dust
component emissivity index, $f_i$ is the fraction of power absorbed and 
re-emitted by dust component $i$ (with $\sum_k f_k = 1$), and $q_i$ is the 
ratio of the FIR emission cross-section to the UV/optical absorption 
cross-section for the $i^{\rm th}$ component (see FDS99 $\S$4 for details). 
We use both the K96 one-component and the FDS99 two-component dust models
in our analyses here.

We need a high resolution Galactic emission map to model foreground 
contamination in the 90 GHz {\it MAP} CMB anisotropy data. Schlegel et al. 
(1998) present a well-calibrated and Fourier-destriped full-sky 100 $\um$ 
Galactic emission map constructed from the {\it COBE}-DIRBE and {\it IRAS}-ISSA 
data sets, with zodiacal light and point source contamination removed. This map
has an angular resolution of $6\farcm1$ FWHM. We smooth this map using a 
Gaussian filter to get a map with total smoothing scale of $0\fdg21$ FWHM,
and estimate the brightness temperature\footnote{
  Brightness temperature $T_A$ is related to intensity $I_{\nu}$ through 
  $I_{\nu}=2kT_A (\nu^2/c^2)$ where $k$ is Boltzmann's constant.}
at 90 GHz by using the Galactic emission spectrum template shown in Figure 1.
For the FDS99 two-component dust model we estimate the 90 GHz dust emission
by extrapolating the 100 $\mu$m flux using the dust temperature and DIRBE
color-correction factor for each line of sight in the maps (see eq. [6] in 
$\S$4 of FDS99). The averaged FDS99 dust emission spectrum (the thick 
long-dashed curve in Fig. 1) is a factor of 1.4 lower than that of the 
K96 dust model. Since the FDS99 model does not include free-free emission,
we generate a Galactic emission contamination map by adding free-free emission,
assumed to be proportional to dust emission and to have an amplitude equal to 18\% of the dust emission amplitude at 90 GHz, which is the percentage of the 
excess dust-correlated DMR emission estimated by FDS99 (see Table 4 of FDS99). 
Thermodynamic temperature fluctuations with respect to the mean CMB temperature
$T_0=2.728$ K (Fixsen et al. 1996) can be obtained from brightness temperature 
fluctuations multipied by the Planck factor $(e^x-1)^2/x^2e^x$, 
where $x=h\nu/kT_0$.
This results in a high resolution all-sky temperature fluctuation map for  
Galactic emission that contributes to the {\it MAP} 90 GHz band data.

While this is the best available map, it is not ideal. First, the K96 model 
parameters are determined from large (DMR) angular scale data while we study 
small-scale microwave temperature fluctuations. Second, the ratio of free-free 
to dust emission may be position dependent, resulting in a position-dependent 
$A_{ff}$ (while we assume that $A_{ff}$ is a constant). Third, if the free-free 
emission power spectrum is flatter than the dust emission power spectrum 
$P(\ell) \propto \ell^{-3}$ in multipole space, the ratio of free-free to dust 
emission will increase with decreasing angular scale (we assume that the large 
scale dust emission correlation with free-free emission persists to smaller 
scales).

In this paper we use the K96 model free-free emission term to account for 
about 20\% of the dust-correlated emission (see FDS99 for a related 
discussion). Similar correlations have been detected for many CMB observations
(see $\S$1, and Kogut 1999 for a review). 
This excess microwave emission is clearly correlated with the dust but is not 
due to thermal (vibrational) dust emission. Draine \& Lazarian (1998) propose 
that K96's dust-correlated emission is not free-free emission but emission 
from rapidly spinning dust grains. Nevertheless, even if the dust-correlated 
emission is not free-free emission, our phenomenological approach is fairly 
reasonable, especially at 90 GHz.

\subsection{Far Infrared Extragalactic Sources}

The {\it IRAS} 1.2 Jy flux-limited galaxy catalog (Strauss et al. 1990;
Fisher et al. 1995) contains 5,319 galaxies, almost all of which are inactive
spiral galaxies though some are active star-forming galaxies and active
galactic nuclei. The IR emission from inactive spiral galaxies is from 
reradiation of interstellar dust heated by the general interstellar 
radiation field (Helou 1986). Assuming that these {\it IRAS} galaxies are 
all inactive spirals with emission spectrum similar to that of our 
Galaxy, we transform their 100 $\um$ fluxes to 90 GHz by using the K96 
Galactic emission spectrum. We also use the method of Gawiser \& Smoot (1997) 
and  Gawiser et al. (1998a) to determine the rms thermodynamic temperature 
fluctuation caused by {\it IRAS} extragalactic sources, and obtain $\rms \simeq
5 \times 10^{-7}$ at 90 GHz and $0\fdg21$ FWHM angular resolution. 
This is consistent with the result of Gawiser \& Smoot (1997).

A second set of contaminating FIR extragalactic sources are the few hundred
sources we have tentatively detected in the DIRBE maps.
DIRBE observed the whole sky with a $0\fdg7 \times 0\fdg7$ square beam and 
produced infrared maps at 10 near-to-far infrared wavelengths, with pixel size 
of $0\fdg32 \times 0\fdg32$. We use the ZSMA (Zodiacal light-Subtracted Mission
Average) maps at 12, 25, 60, 100, 140, and 240 $\um$ to search for dim signals 
in the maps. We generate a difference map at each band by subtracting the 
intensity map smoothed over $2\fdg36$ FWHM from the original unsmoothed map. 
Thus only signals on scales smaller than $2\fdg36$ remain in the difference 
maps. The $i^{\rm th}$ pixel noise $\sigma_i$ in the difference map is 
$\sigma_i = \sigma_N / \sqrt{N_i}$, where $\sigma_N$ is the global noise level 
of the difference map and $N_i$ is the number of observations 
at the $i^{\rm th}$ pixel. 
$\sigma_N$ can be determined iteratively from the less-contaminated background 
region ($|b|>60\deg$ and ecliptic latitude $|\beta|>20\deg$) by requiring $z = 
I_i / \sigma_i$ (where $I_i$ is the intensity at $i^{\rm th}$ pixel) to have a 
normal distribution with zero mean and unit variance at $z<0$. The resulting 
global noise levels are 0.209, 0.295, 0.549, 1.14, and 34.8, and 19.9 MJy/sr 
at 12, 25, 60, 100, 140, and 240 $\um$, respectively. The noise levels in the 
140 and 240 $\um$ difference maps are very large so we exclude these maps from
our analysis.

In the region $|b|>20\deg$ of the 100 $\um$ map we select pixels with
$I_{i,100~\um} > 3\sigma_i$. A faint extragalactic source may appear in at
most 4 pixels ($0\fdg64 \times 0\fdg64$) in the DIRBE map, when it is located 
where the 4 pixels touch. By grouping contiguous pixels, we search for signals 
in groups of 4 or fewer pixels in the Quadrilateralized Spherical Cube 
projection (see {\it COBE} DIRBE Exp. Supp. 1998). 
This excludes most of the bright galaxies detected by Odenwald, Newmark, \& 
Smoot (1998), since they are typically detected in more than 4 pixels. 
We find 421 and 401 such groups (with 4 or fewer pixels) in the Galactic north 
and south hemispheres, respectively. 436 of these groups have positive fluxes 
at all wavelengths considered. We finally catalogue 289 extragalactic source 
candidates satisfying the {\it IRAS} bright galaxy color criteria (Smith et al.
1987; Soifer et al. 1989; Strauss et al. 1990). These color criteria are
$-0.9 < \log(f_{60~\um} / f_{100~\um} ) < 0.3$, 
$-1.5 < \log(f_{25~\um} / f_{60~\um} ) < 0.0$,
$-1.2 < \log(f_{12~\um} / f_{25~\um} ) < 0.4$,
$-2.1 < \log(f_{12~\um} / f_{60~\um} ) < 0.0$, and
$f_{60~\um}^2 > f_{12~\um} f_{25~\um}$.
Figure 2 shows the distribution of the final 289 DIRBE FIR sources. 33 
{\it IRAS} galaxies and only one source in the 5 GHz 1 Jy flux-limited 
extragalactic radio source catalog (518 objects, K\"uhr et al. 1981) are within
a radius of $0\fdg32$ of these DIRBE sources. Most of these sources are matched 
with extragalactic objects in the {\it IRAS} Faint Source Catalog within the 
same radius. However, the large DIRBE pixel size prevents us from uniquely
identifying these sources as extragalactic because each DIRBE pixel contains
many {\it IRAS} sources, some of which are not extragalactic.
Some of the DIRBE sources could conceivably be objects in our Galaxy that 
coincidentally satisfy the {\it IRAS} galaxy color criteria. The emission from 
these DIRBE FIR sources (after exclusion of the 33 sources identified with 
{\it IRAS} galaxies) are transformed to 90 GHz thermodynamic temperature 
fluctuations using the method applied above to the {\it IRAS} galaxies.

\subsection{Radio Point Sources}

The 4.85 GHz Green Bank survey catalog (Becker, White, \& Edwards 1991)
contains 53,522 radio point sources with fluxes larger than about 40 mJy and
covers $0\deg < \delta < 75\deg$. The 4.85 GHz Parkes-MIT-NRAO survey
catalog (Griffith \& Wright 1993) has 50,814 point sources with a flux limit
of about 35 mJy in the $-87\fdg5 < \delta < 10\deg$ region. Besides these, 
there is a 1.4 GHz survey catalog that has 30,239 sources flux-limited at 100 
mJy in the $-5\deg < \delta < 82\deg$ region (White \& Becker 1992).
These catalogs contain extragalactic sources as well as Galactic radio sources.
For example, the Green Bank survey catalog has about 350 galaxies, $\sim 800$
quasars, more than 500 BL Lacertae objects, and $\sim 400$ X-ray sources
(Becker et al. 1991).

To make a combined radio source catalog that almost covers the sky, we 
adopt the Green Bank catalog for $0\deg < \delta < 75\deg$, the Parkes-MIT-NRAO 
catalog for $-87\fdg5 \le \delta \le 0\deg$, and the 1.4 GHz radio source 
catalog for $75\deg < \delta < 82\deg$. In the few GHz frequency band, the main 
emission mechanism for radio sources is synchrotron radiation with power-law 
spectral behavior $I_\nu \sim \nu^{\alpha}$ with $-1.0 \lesssim \alpha 
\lesssim -0.5$ (Platania et al. 1998). Since spectral indices of most sources 
in the Green Bank and Parkes-MIT-NRAO catalogs are close to $-0.85$, radio 
sources with unknown $\alpha$ are assumed to have $\alpha = -0.85$. Some radio 
sources with compact active nuclei generate flat-spectrum radio emission 
($\alpha \ge 0$) due to self-absorption of low energy photons. The average 
spectral index of flat-spectrum compact sources is $\alpha \simeq 0$ at 
$\nu \le 200$ GHz, steepening to $\alpha = -0.7$ at higher frequencies 
(Danese et al. 1987; Toffolatti et al. 1998). In our analyses here we set 
$\alpha=0$ for all radio sources with spectral index $\alpha > 0$. Emission 
from radio sources in the region $75\deg < \delta < 82\deg$ are transformed 
to 4.85 GHz fluxes using the spectral index determined above. The fluxes of 
the final 72,132 4.85 GHz sources, covering $-87\fdg5 < \delta < 82\deg$, 
with flux limit of 43 mJy, are converted to temperature fluctuations when 
``observed" with the $0\fdg21$ FWHM {\it MAP} beam at 90 GHz.

\section{Generating Mock {\it MAP} Maps}

Park et al. (1998, hereafter P98) generate mock CMB anisotropy maps for the 
90 GHz ($0\fdg21$ FWHM beamwidth) channel of the {\it MAP} experiment in a 
spatially-flat $\Lambda$ dominated CDM model. This $\Lambda$CDM cosmogony 
has non-relativistic matter density parameter $\omz = 0.4$, $h=0.6$ (where the 
Hubble constant $H_0 = 100 h$ km s$^{-1}$ Mpc$^{-1}$), baryonic-matter 
density parameter $\om_B = 0.0125 h^{-2}$, and is normalized to the four-year 
DMR data (Ratra et al. 1997). The rms level of the CMB anisotropy in 
this map is 104 $\uK$ (at $0\fdg21$ FWHM), larger than that of the flat
CDM case (92 $\uK$; Refregier, Spergel, \& Herbig 2000). The maps include
{\it MAP} instrumental noise, simulated to give 35 $\uK$ per $0\fdg3 \times 
0\fdg3$ pixel.

It is expected that {\it MAP} will produce CMB temperature fluctuation maps 
with zero mean even if foreground contamination is not completely removed.
Thus in the map the observed temperature at each position on the sky,
$T(\theta,\phi)$, is the sum of the CMB contribution $T_0 + \delta T_{CMB}$, 
a contribution from foreground contamination $\delta T_C$, and an instrumental 
noise contribution $\delta T_N$, where $T_0$ is CMB black body temperature 
(2.728 K) and $\delta T_{CMB}$ is the CMB anisotropy temperature.
Therefore the observed temperature anisotropy is
$$ 
   \tot(\theta,\phi)
   \equiv { {T(\theta,\phi)-\langle T \rangle} \over {\langle T \rangle} }
        \simeq { {\delta T_{CMB} (\theta,\phi) + \delta T_C (\theta,\phi)
        + \delta T_N (\theta,\phi) - \langle \delta T_C \rangle}
        \over {T_0} }  
   \eqno (2) 
$$
where $\langle \cdots \rangle$ represents an average over the whole sky 
excluding  the Galactic plane region (typically the $|b|>20\deg$ part of 
the sky in our analyses here), and $\langle \delta T_{CMB} \rangle = \langle 
\delta T_N \rangle = 0$ are assumed. The term $\langle \delta T_C \rangle$
that will appear in the denominator of the last expression is omitted, since 
it is negligible compared with $T_0$.

We use eq. (2) to generate foreground-contaminated CMB temperature 
fluctuation maps at 90 GHz with $0\fdg21$ FWHM resolution. 
These maps are further smoothed over $0\fdg214$ (i.e., a total smoothing 
scale of $0\fdg3$ FWHM) to reduce noise, and consist of $3600 \times 1801$ 
points in $(\ell,b)$ space with $0\fdg1$ spacing. Figure 3 shows one of the 
foreground-contaminated CMB temperature fluctuation maps (using the K96 
Galactic emission model). The Galactic emission model has an uncertainty. 
Its uncertainty (hereafter $\Delta$G) is taken to be the difference between 
the maps generated from the K96 (constant dust temperature) and FDS99 
(spatially-varying dust temperature) Galactic emission models.
Table 1 lists the map rms values of the CMB temperature fluctuations, {\it MAP}
instrumental noise, and various foreground emissions.
Note that rejecting radio sources above 5 $\sigma$ reduces the rms level 
of temperature fluctuations. When analyzing the maps we mask the LMC, SMC, M31 
and Orion-Taurus regions for Galactic emission-contaminated maps. 
For radio source emission-contaminated maps we mask regions that are not 
surveyed ($\delta > 82\deg$, $\delta < -87.5\deg$) or only partially surveyed.

\section{Correlation Function}

\subsection{Amplitude of Correlation Function}

The two-point angular correlation function (hereafter CF) for temperature 
anisotropy is $C(\theta) \equiv \langle \delta T/T(\theta_i,\phi_i)
\delta T/T(\theta_j,\phi_j) \rangle$, where $\langle \cdots \rangle$ denotes
an ensemble average and $\theta$ is the angular separation between 
$(\theta_i,\phi_i)$ and $(\theta_j,\phi_j)$. 
It is related to the angular power spectrum $C_\ell$ through
$$ 
     C(\theta) = \sum_{\ell=2}^{\infty} {{2\ell+1} \over {4\pi}} C_{\ell}
               P_{\ell} (\cos \theta) B_{\ell}^2 F_{\ell}^2,
     \eqno (3)
$$
where $P_\ell$ is a Legendre polynomial and $B_\ell$ and $F_\ell$ are the 
coefficients of the beam function and smoothing filter expanded in a Legendre
polynomial basis.

We measure $C(\theta)$ directly from the CMB maps by picking up a million 
pairs of points for each angular bin considered.
Figure 4$a$ shows the angular CFs of noise-free and noise-added $\Lambda$CDM 
CMB anisotropy maps. The theoretical CMB anisotropy CF of the $\Lambda$CDM 
cosmological model (dot-dashed line) is also shown for comparison. Note that 
the amplitude of the CF can exhibit large cosmic variance while the position of 
the acoustic valley does not (see $\S$3 of P98 for the effect of cosmic 
variance on the position of the acoustic valley of the CF). The amplitude of 
the CF of the noise-added CMB map is slightly higher than that of the 
noise-free case on scales less than $0\fdg3$. The effect of noise on the CF 
amplitude exceeds that of cosmic variance on these scales. On scales larger 
than $0\fdg3$, however, instrumental noise hardly affects the CF and cosmic 
variance dominates the statistical uncertainty. The effect on the CF of 
Galactic plane cuts at $|b|=20\deg$ and $40\deg$ is illustrated in Figure 4$b$. 

The effects of foreground emission on the CF are shown in Figure 5, where
the CFs are computed for the Galactic plane cut region with $|b|>20\deg$.
Although foreground Galactic emission, modeled either by K96 or FDS99, 
slightly increases the correlation amplitude on all angular scales (Fig. 
5$a$), these effects are smaller than the cosmic and sample variance 
uncertainties. 
Figure 5$b$ demonstrates that the Galactic emission model uncertainty 
$\Delta$G (i.e., Galactic emission residuals which remain after removing 
model Galactic emission from a foreground-contaminated map) hardly
affects the CF. {\it IRAS} and DIRBE FIR extragalactic sources appear not 
to have any effect (Fig. 5$c$). Radio point source emission raises the 
CF amplitude on scales smaller than $0\fdg5$. When bright radio sources above 
5 $\sigma$ are removed this small-scale effect becomes much weaker (Fig. 5$d$).

\subsection{Position of the Acoustic Valley}

The Doppler peaks in the angular power spectrum, particularly the first Doppler 
peak, appear as an acoustic valley in $\theta^{1/2} C(\theta)$. The position, 
depth, and width of the acoustic valley depend on the cosmological model 
parameters (P98). In addition to the effect of foreground contamination on the 
amplitude of the CF, it is necessary to study their effect on the position of 
the acoustic valley.

To predict acoustic valley positions we compute $C(\theta)$ (using eq. [3]) 
from 200 sets of $a_{\ell}^{m} B_{\ell}$ and $(a_{\ell}^{m} B_{\ell} + 
n)F_{\ell}$ up to $\ell = 2500$. We assume the noise $n$ is a Gaussian random 
variable with zero mean and variance appropriate for {\it MAP}, and here 
$B_\ell$ and $F_\ell$ represent the $0\fdg21$ FWHM beam and $0\fdg214$ FWHM 
smoothing filter, respectively. The uncertainty in the position of the acoustic
valley was estimated from these 200 Monte Carlo realizations (P98).
The $\Lambda$CDM model predicts the location of the acoustic valley to be at
$\theta_{\rm v}=1\fdg12\pm0\fdg02$ for a $0\fdg21$ beam. When we account for 
the {\it MAP} noise and use an additional $0\fdg214$ FWHM smoothing filter 
(total smoothing of $0\fdg3$ FWHM), the position becomes $\theta_{\rm v} = 
1\fdg14\pm0\fdg02$ while it is $0\fdg65\pm0\fdg04$ for an open CDM model
with $\Omega_0 = 0.4$, $h=0.65$. 

We have also computed $C(\theta)$ near the acoustic valley directly from the 
foreground-contaminated CMB anisotropy maps, and have measured the position 
of the acoustic valley. The results are shown in Figure 6 and Table 2.
The location of the acoustic valley is quite robust. When foreground Galactic 
emission is reasonably modeled, the Galactic cut mainly affects its location
($\sim 0\fdg01$). The cosmic variance is about $0\fdg02$. 
  
\section{Topology} 

\subsection{The Genus}

Gaussianity and phase correlation information are
important probes of the origin of structure formation. Topology measures for 
two-dimensional fields, first introduced by Coles \& Barrow (1987), Melott et 
al. (1989) and Gott et al. (1990), have been applied to observational data 
such as the angular distribution of galaxies on the sky (Coles \& Plionis 1991; 
Gott et al. 1992), the distribution of galaxies in slices of the universe 
(Park et al. 1992; Colley 1997; Park, Gott, \& Choi 2001a), and the CMB
temperature fluctuation field (Coles 1988; Gott et al. 1990; Smoot et al. 1994; 
Kogut et al. 1996c; Colley, Gott, \& Park 1996; Park et al. 2001b).
 
We use the two-dimensional genus statistic introduced by Gott et al. (1990)
as a quantitative measure of the topology of the CMB anisotropy. 
For the CMB the genus is the number of hot spots 
minus the number of cold spots. Equivalently, the genus at a temperature 
threshold level $\nu$ is
$$
    g(\nu) = {1 \over 2\pi} \int_C \kappa ds,
    \eqno (4)
$$
where $\kappa$ is the signed curvature of the iso-temperature contours $C$,
and the threshold level $\nu$ is the number of standard deviations from
the mean. At a given threshold level, we measure the genus by integrating 
the curvature along iso-temperature contours. The curvature is positive 
(negative) if the interior of a contour has higher (lower) temperature than 
the specified threshold level.

The genus of a two-dimensional random-phase Gaussian field is $g(\nu) \propto 
\nu e^{-\nu^2/2}$ (Gott et al. 1990). Non-Gaussian features in the CMB 
anisotropy will appear as 
deviations of the genus curve from this relation (P98; Park et al. 2001a, 
2001b). Non-Gaussianity can shift the observed genus curve to the left
(toward negative thresholds) or right near the mean threshold level.
The shift $\Delta \nu$ of the genus curve with respect to the Gaussian relation
is measured by minimizing the $\chi^2$ between the data and the fitting function
$$
    G = A_s \nu^\prime e^{-{\nu^\prime}^2 /2},
    \eqno (5)
$$
where $\nu^\prime = \nu - \Delta\nu$, and the fitting is performed over the 
range $-1.0 \le \nu \le 1.0$.
Non-Gaussianity can also alter the amplitudes of the genus curve at positive 
and negative levels differently causing 
$|g(\nu \approx -1)| \neq |g(\nu \approx +1)|$.
The asymmetry parameter is defined as
$$
    \Delta g = A_H - A_C,
    \eqno (6)
$$
where
$$
    A_H = \int_{\nu_1}^{\nu_2} g_{\rm obs} d\nu
         / \int_{\nu_1}^{\nu_2} g_{\rm fit} d\nu,
    \eqno (7)
$$
and likewise for $A_C$. The integration is limited to $-2.4 \le \nu \le -0.4$
for $A_C$ and to $0.4 \le \nu \le 2.4$ for $A_H$. The overall amplitude $A$ of
the best-fit Gaussian genus curve $g_{\rm fit}$ is found from $\chi^2$-fitting
over the range $-2.4 \le \nu \le 2.4$. Positive $\Delta g$ means that more hot
spots are present than cold spots. This kind of topological information 
can not be drawn simply from the one-point distribution. 
In summary, for a given genus curve we measure the best-fit amplitude $A$, 
the shift parameter $\Delta\nu$, and the asymmetry parameter $\Delta g$.

\subsection{Effects of Foregrounds on CMB Anisotropy Topology}

We use our mock foreground-contaminated CMB anisotropy maps for the {\it MAP} 
experiment to study the sensitivity of the genus statistic to non-Gaussianity 
of cosmological origin. To compute the genus we stereographically project the 
maps on to a plane. This conformal mapping locally preserves shapes of 
structures. Additional Gaussian smoothings are applied during the projection 
so that the projected maps have total smoothing scales of $0\fdg3$, $0\fdg5$ 
and $1\fdg0$ FWHM. We then exclude regions with $|b|<20\deg$ to reduce the 
effect of Galactic emission. We present the genus curves as a function of 
the area fraction threshold level $\nu_A$, which is defined to be the 
temperature threshold level at which the corresponding iso-temperature contour 
encloses a fraction of the survey area equal to that at the temperature 
threshold level $\nu_A$ for a Gaussian field (Gott et al. 1990). 
The $\nu_A = 0$ level corresponds to the median temperature because this
threshold level divides the map into high and low regions of equal area.
All our analyses are based on genus curves measured as a function
of $\nu_A$.

Figure 7 shows genus curves measured from the mock {\it MAP} CMB maps 
in the $\Lambda$CDM model with various foreground contaminations at $0\fdg3$ 
FWHM resolution (open symbols). The genus curve of the foreground-free map 
(noise-added; filled circles) is shown for comparison. 
The effect of instrumental noise on the genus is significant on both $0\fdg3$ 
and $0\fdg5$ scales (filled circles in Fig. 7$a$; also see the genus amplitude 
parameter $A$ values in Table 3 described below). 
It should also be noted that the $0\fdg3$ scale genus is reduced by
the finite pixel size of the {\it MAP} simulation (Melott et al. 1989). 

Table 3 lists the genus-related statistics $A$, $\Delta\nu$, and $\Delta g$
measured from CMB maps with $0\fdg3$, $0\fdg5$, and $1\fdg0$ FWHM resolutions
using eqs. (5) to (7). The uncertainties are $1\sigma$ errors derived from 
the variances over the eight octants and account for cosmic variance. Here 
$\Delta\nu_0$ and $\Delta g_0$ are the values from the noise- and 
foreground-free pure CMB anisotropy maps. Hence $\Delta\nu-\Delta\nu_0$ and 
$\Delta g - \Delta g_0$ are the net effects due to instrumental noise or 
foreground emission. Table 3 shows that Galactic emission significantly affects 
the shift parameter on all angular scales, at a level exceeding 2 $\sigma$.
However, their effect become much less serious when only the Galactic emission
model uncertainty $\Delta$G needs to be accounted for. 
FIR extragalactic source emission has minimal effect on the genus, on all 
angular scales. Radio point source emission results in positive genus asymmetry,
particularly on intermediate angular scales ($0\fdg5$ FWHM) at a statistical
significance level of about 2 $\sigma$. We have also measured the genus from 
the radio source-contaminated CMB anisotropy map with $0\fdg5$ FWHM resolution 
after broadly excising pixels above 4.7 times the rms level of the noise-added 
CMB anisotropy map, and find the asymmetry still exceeds cosmic variance 
($\Delta g - \Delta g_0 = 0.035\pm0.020$). 

When Galactic diffuse emission and bright radio point source emission are 
removed from the observed temperature fluctuation map using accurate foreground
models, one may conclude that non-Gaussianity of cosmological origin is present
if the measured genus-related parameters $|\Delta \nu| \gtrsim 0.02$ ($0.04$) 
or $|\Delta g| \gtrsim 0.03$ ($0.08$) at $0\fdg3$ ($1\fdg0$) FWHM scale at 
2 $\sigma$.

\section{Conclusions}

We study the effects of foreground contamination and instrumental noise on the
simulated {\it MAP} data through the correlation function and genus
statistics. The foregrounds we consider are diffuse Galactic dust and
free-free emissions, far infrared extragalactic sources, and radio point 
sources. The mock observations are made in a spatially-flat $\Lambda$CDM model.

Our study shows that FIR extragalactic sources in the {\it IRAS} 1.2 Jy 
galaxy catalog and the {\it COBE}-DIRBE map have negligible effect on the
correlation function and the genus. Radio point sources also have little effect 
on the correlation function when sources with amplitudes larger than 5 times 
the rms of the noise-added $\Lambda$CDM CMB anisotropy map are removed. 
However, on $0\fdg5$ scales radio sources cause positive genus curve asymmetry, 
a non-Gaussian feature, at a statistical significance level of about 2 $\sigma$ 
compared to cosmic variance (Table 3). This is the scale where the smoothing 
scale is small enough that strong radio sources are resolved, and where it is 
large enough that instrumental noise is not a dominant anisotropy source. 
This non-Gaussian feature in the genus curve is not significantly reduced by 
removing sources above 5 $\sigma$. We conclude that good radio point source 
data at 90 GHz is essential for an unambiguous detection of non-Gaussian 
behavior of the CMB anisotropy on this scale.

Diffuse Galactic emission affects the correlation function much less than 
cosmic variance. But it can produce negative shifts of the genus curve, 
another non-Gaussian feature, which are 2 or 3 times larger than cosmic
variance. These effects are significantly reduced if model Galactic emissions  
(e.g., K96 or FDS99) are subtracted from the observed {\it MAP} data, unless 
the amplitude of Galactic foreground emission in these models are in great 
error. The typical shift of the genus curve due to the uncertainty in the 
Galactic emission models is estimated from the difference map between the
K96 and FDS99 models. On scales below $1\fdg0$ it is of order of the shift 
due to cosmic variance. Therefore, an accurate Galactic emission model is 
still necessary to make reliable tests of Gaussianity of the CMB anisotropy. 
With the present understanding of Galactic emission and radio sources, one will 
be able to detect non-Gaussianity of cosmological origin at the 2 $\sigma$ 
level if the shift parameter of the genus curve $|\Delta\nu| \gtrsim 0.02$ 
($0.04$), or the asymmetry parameter $|\Delta g| \gtrsim 0.03 $ ($0.08$) on $0\fdg3$ ($1\fdg0$) FWHM scales.

There are other kinds of foreground contamination emission mechanisms,
in addition to those discussed above, such as the thermal Sunyaev-Zel'dovich 
effect, the Ostriker-Vishniac effect, gravitational lensing, and so on 
(see Refregier et al. 2000 for a review). Some of these will need to be dealt 
with if we are to fully utilize the cosmological information in the anticipated
{\it MAP} and {\it Planck} data.


We acknowledge valuable discussions with D. Finkbeiner, K. Ganga, Juhan Kim, 
Sungho Lee, and P. Mukherjee, and helpful comments from the anonymous referee. 
CGP and CP acknowledge support from the BK21 
program of the Korean Government and the Basic Research Program of the Korea 
Science \& Engineering Foundation (grant no. 1999-2-113-001-5). BR acknowledges 
support from NSF CAREER grant AST-9875031. We also acknowledge support
from NSF grant AST-9900772 to J.R. Gott.

\clearpage

\begin{deluxetable}{lc}
\tablewidth{0pt}
\tablecaption{Contributions to Mock 90 GHz {\it MAP} Data at 
              $0\fdg3$ FWHM Resolution}  
\tablehead{
\colhead{Maps} &
\colhead{RMS ($\uK$)}
}
\startdata
CMB                       & $100.3$  \\
{\it MAP} noise           & $35.8$  \\
CMB+noise                 & $106.5$ \\
Galactic emission (K96)   & $14.0$ \\
Galactic emission (FDS99) & $12.5$ \\
$\Delta$G (K96-FDS99)     & $6.1$ \\ 
Radio sources             & $26.8$  \\
Radio sources ($>5\ \sigma$ rejected) & $16.7$ \\
FIR extragalactic sources & $2.0$ \\
\enddata
\end{deluxetable}

\begin{deluxetable}{llc}
\tablewidth{0pt}
\tablecaption{Correlation Function Acoustic Valley Position Measured
              from Mock $\Lambda$CDM Skies}
\tablehead{
\colhead{} &
\colhead{Map} &
\colhead{$\theta_{\rm v}$}
}
\startdata
No cut       & CMB      & $1\fdg122$    \\
$$           & ~~+noise & $1\fdg124$    \\
\hline
$|b|>20\deg$ & ~~+noise & $1\fdg113$    \\
$$  & ~~+noise+Galactic emission (K96)    & $1\fdg109$  \\
$$  & ~~+noise+Galactic emission (FDS99)  & $1\fdg109$  \\
$$  & ~~+noise+$\Delta$G                  & $1\fdg114$  \\
$$  & ~~+noise+FIR extragalactic sources  & $1\fdg114$  \\
$$  & ~~+noise+radio sources              & $1\fdg112$  \\
\enddata
\end{deluxetable}

\begin{deluxetable}{llrrr}
\tablewidth{0pt}
\tablecaption{Genus Statistics Measured from Mock {\it MAP} Data}
\tablehead{
\colhead{Res.}  &
\colhead{Maps}  &
\colhead{$A$}  &
\colhead{$\Delta\nu - \Delta\nu_0$}  &
\colhead{$\Delta g - \Delta g_0$}  
}
\startdata
$0\fdg3$ & CMB      & 
   $961\pm23$   & $ 0.000\pm0.007$ & $0.000\pm0.011$\\ 
$$       & ~~+noise & 
   $1746\pm36$  & $ 0.001\pm0.009$ & $-0.008\pm0.013$\\
$$       & ~~+noise+Galactic emission (K96) & 
   $1730\pm32$  & ${\bf -0.017\pm0.009}$ & $-0.017\pm0.015$\\
$$       & ~~+noise+Galactic emission (FDS99) &
   $1734\pm33$  & ${\bf -0.024\pm0.009}$ & $-0.017\pm0.014$ \\
$$       & ~~+noise+$\Delta$G &
   $1739\pm31$  & $ 0.006\pm0.008$ & $-0.003\pm0.013$ \\
$$       & ~~+noise+radio sources & 
   $1775\pm123$ & $-0.001\pm0.010$ & $0.003\pm0.013$ \\
$$       & ~~+noise+FIR extragalactic sources & 
   $1747\pm36$  & $0.002\pm0.009$  & $-0.007\pm0.013$ \\
\hline
$0\fdg5$ & CMB      & 
    $553\pm18$ & $0.000\pm0.012$ & $0.000\pm0.020$ \\ 
$$       & ~~+noise  & 
    $614\pm18$ & $0.001\pm0.012$ & $0.003\pm0.021$ \\
$$       & ~~+noise+Galactic emission (K96) & 
    $605\pm15$ & ${\bf -0.025\pm0.013}$ & $0.000\pm0.024$ \\
$$       & ~~+noise+Galactic emission (FDS99) &
    $606\pm17$ & ${\bf -0.034\pm0.013}$ & $0.003\pm0.022$ \\
$$       & ~~+noise+$\Delta$G & 
    $612\pm16$ & $ 0.010\pm0.012$ & $ 0.002\pm0.022$ \\
$$       & ~~+noise+radio sources & 
    $629\pm49$ & $-0.011\pm0.017$ & ${\bf 0.041\pm0.020}$ \\
$$       & ~~+noise+FIR extragalactic sources & 
    $615\pm18$ & $0.000\pm0.012$ & $0.004\pm0.021$ \\
\hline
$1\fdg0$ & CMB      & 
   $181\pm10$ & $0.000\pm0.019$ & $0.000\pm0.036$ \\ 
$$       & ~~+noise  & 
   $184\pm10$ & $0.001\pm0.019$ & $0.003\pm0.037$ \\
$$       & ~~+noise+Galactic emission (K96) & 
   $180\pm8$ & ${\bf -0.033\pm0.020}$ & $0.008\pm0.041$ \\
$$       & ~~+noise+Galactic emission (FDS99) &
   $182\pm9$ & ${\bf -0.044\pm0.019}$ & $0.001\pm0.040$ \\
$$       & ~~+noise+$\Delta$G & 
   $183\pm9$ & $ 0.010\pm0.019$ & $ 0.025\pm0.039$ \\
$$       & ~~+noise+radio sources & 
   $189\pm18$ & $-0.004\pm0.029$ & $0.034\pm0.045$ \\
$$       & ~~+noise+FIR extragalactic sources & 
   $184\pm10$ & $0.002\pm0.019$ & $0.004\pm0.037$ \\
\enddata
\end{deluxetable}

\clearpage

\begin{figure}
\resizebox{\textwidth}{!}{\includegraphics{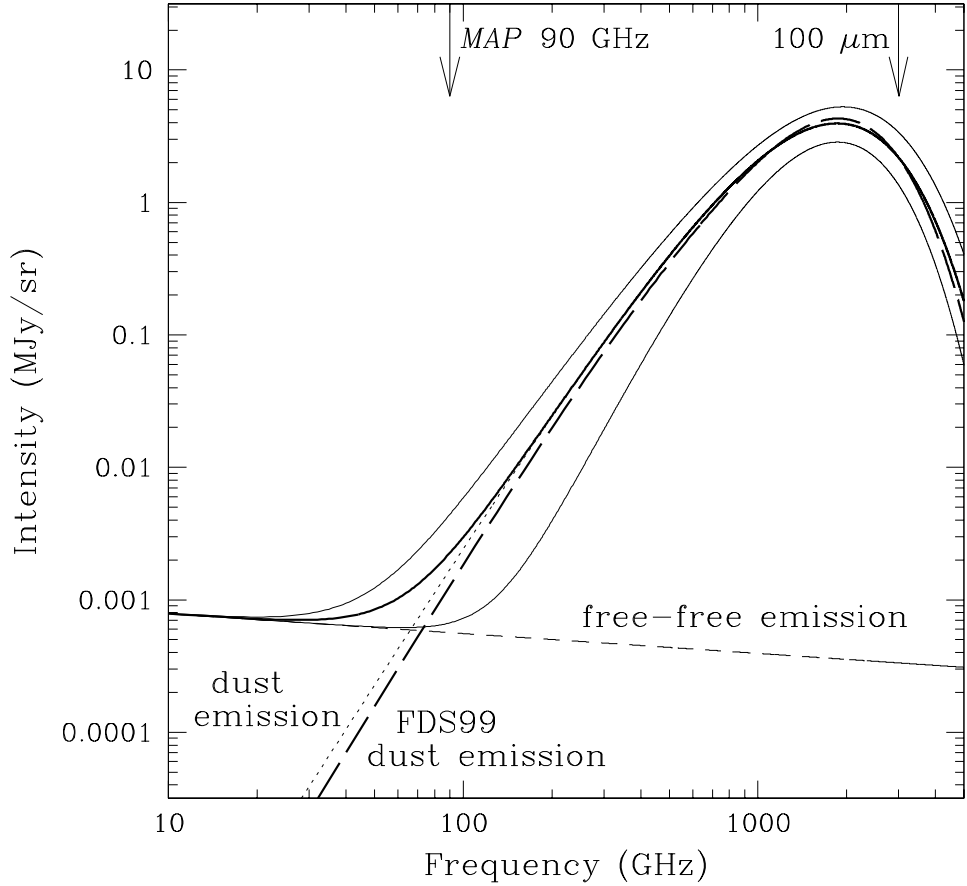}}
\caption{FIR-to-microwave Galactic emission spectra at $|b|>20\deg$.
         K96 Galactic emission spectrum (thick solid curve) is the sum of dust 
         (dotted curve) and free-free (dashed curve) emission components, 
         inferred by cross-correlating the DMR and DIRBE maps. Upper and lower 
         thin solid curves are upper and lower limits of the K96 emission 
         model. The thick long-dashed curve is the FDS99 dust spectrum, derived 
         by averaging all the temperature-varying dust spectra at each line of 
         sight and normalized with K96 spectrum at 3000 GHz (100 $\um$) 
         for direct comparison.}
\end{figure}
\clearpage

\begin{figure}
\resizebox{\textwidth}{!}{\includegraphics{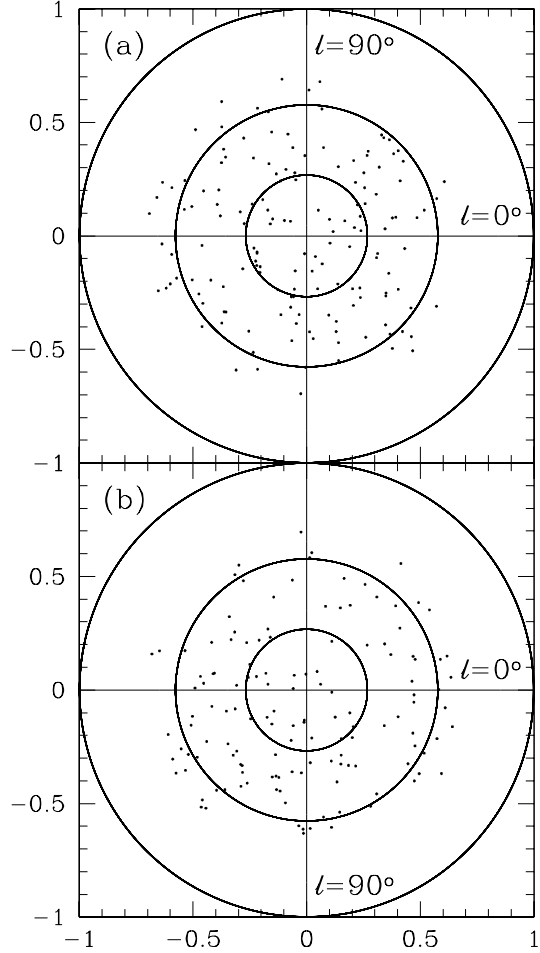}}
\caption{Stereographic projections of 289 DIRBE FIR sources found 
         in ($a$) north (152 groups) and ($b$) south (137 groups) hemispheres.
         The $x$ and $y$ coordinates are defined by $x=\cos l\tan(\pi/2-b)/2$,
         $y=\sin l \tan(\pi/2-b)/2$ for the north hemisphere,
         $x=\cos l / \tan(\pi/2-b)/2$, $y=-\sin l / \tan(\pi/2-b)/2$
         for the south hemisphere. The three circles in each panel, from outside
         to the center, indicate $|b|=0\deg$, $30\deg$, $60\deg$, respectively. 
         The Galactic center is located at $(x,y)=(1,0)$ in both panels.} 
\end{figure}
\clearpage

\begin{figure}
\resizebox{\textwidth}{!}{\includegraphics{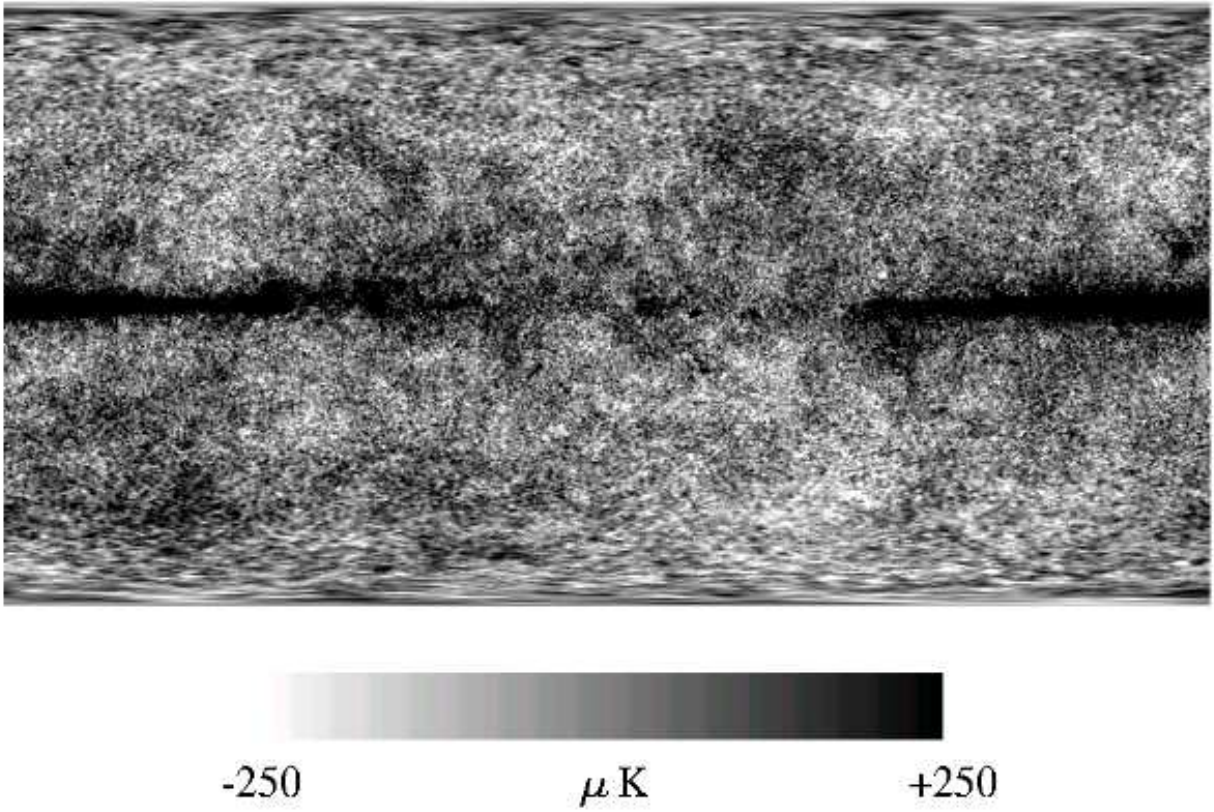}}
\caption{The $\Lambda$CDM model foreground-contaminated CMB anisotropy 
         temperature fluctuation mock map for the 90 GHz {\it MAP} channel,
         smoothed to $0\fdg3$ FWHM angular resolution. Foreground contamination
         here is modelled by the K96 Galactic emission model.}
\end{figure}
\clearpage

\begin{figure}
\resizebox{\textwidth}{!}{\includegraphics{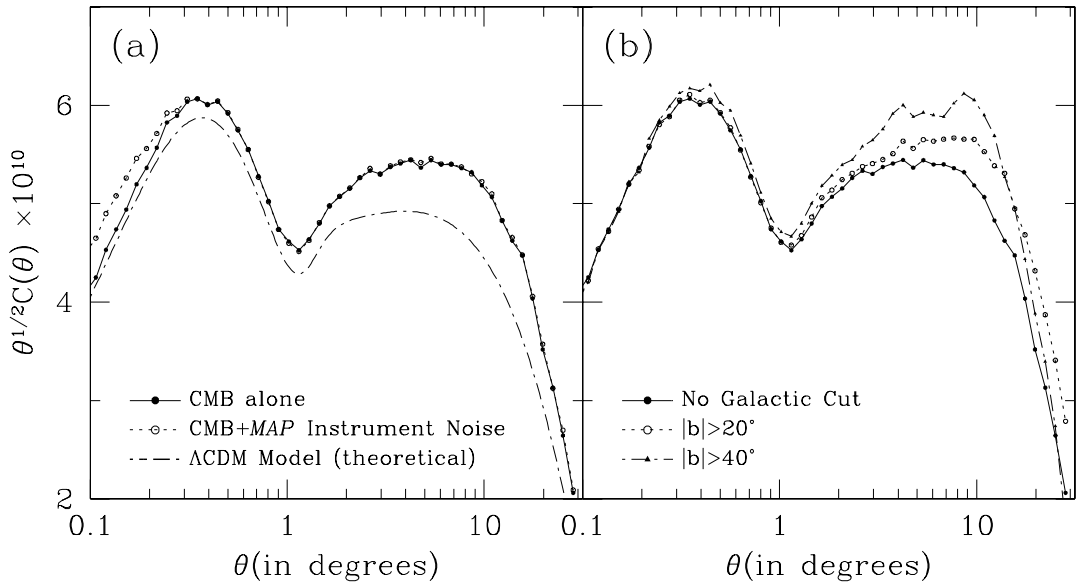}}
\caption{($a$) Two-point angular correlation functions measured from the 
         noise-free (filled circles, solid curve) and noise-added (open circles,
         dotted curve) $\Lambda$CDM temperature anisotropy maps at $0\fdg3$ 
         FWHM resolution. The theoretical CMB anisotropy angular correlation 
         function for the $\Lambda$CDM model (dot-dashed curve) is shown for 
         comparison. ($b$) Noise-added $\Lambda$CDM temperature two-point 
         angular correlation functions showing the effect of different 
         Galactic plane cuts. For $|b|>40\deg$ (filled triangles, dot-dashed 
         curve), the angular correlation appears to be strongly biased by 
         sample variance.}
\end{figure}
\clearpage 

\begin{figure}
\resizebox{\textwidth}{!}{\includegraphics{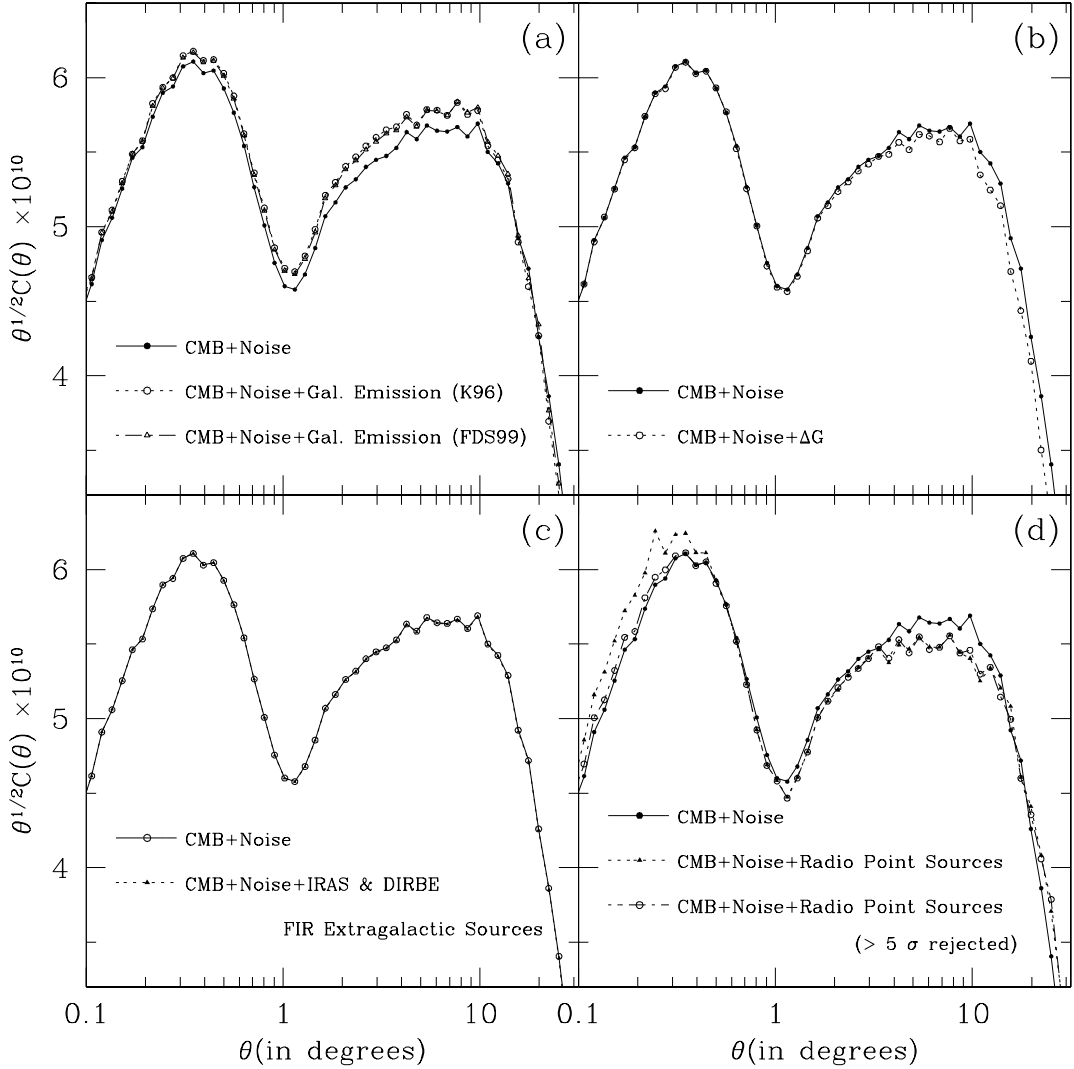}}
\caption{Two-point angular correlation functions measured from the noise-added
         $\Lambda$CDM model CMB anisotropy maps contaminated by ($a$) 
         Galactic emission, ($b$) the Galactic emission model uncertainty, 
         ($c$) FIR extragalactic source emission, and ($d$) radio point source
         emission. All correlation functions are measured from the Galactic 
         plane cut maps ($|b|>20\deg$). In all panels, the foreground-free 
         correlation function (solid curve) measured from the noise-added 
         $\Lambda$CDM model map is shown for comparison. Note that 5 $\sigma$ 
         rejection of bright radio point sources effectively restores the 
         radio source-free result ($d$, open circles, dot-dashed curve).} 
\end{figure}
\clearpage 

\begin{figure}
\resizebox{\textwidth}{!}{\includegraphics{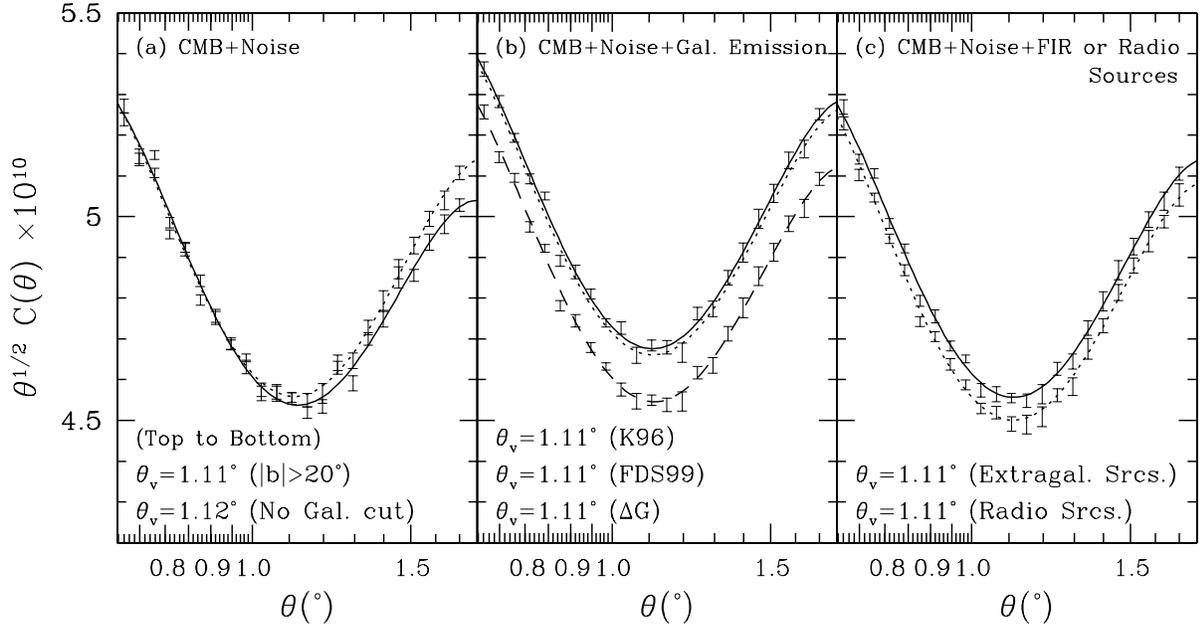}}
\caption{Angular correlation functions in the region of acoustic valleys. Error 
         bars are $1\sigma$ uncertainties estimated from 15 independent 
         correlation function measurements with different random seeds to pick 
         up a million pairs of points differently. 
         Solid, dotted, and dashed curves are 4$^{\rm th}$-order polynomial 
         fits used to measure acoustic valley positions, listed as 
         $\theta_{\rm v}$ in the lower-left part of each panel. 
         Error bars for FDS99 case in ($b$) are omitted for simplicity.}
\end{figure}
\clearpage 

\begin{figure}
\resizebox{\textwidth}{!}{\includegraphics{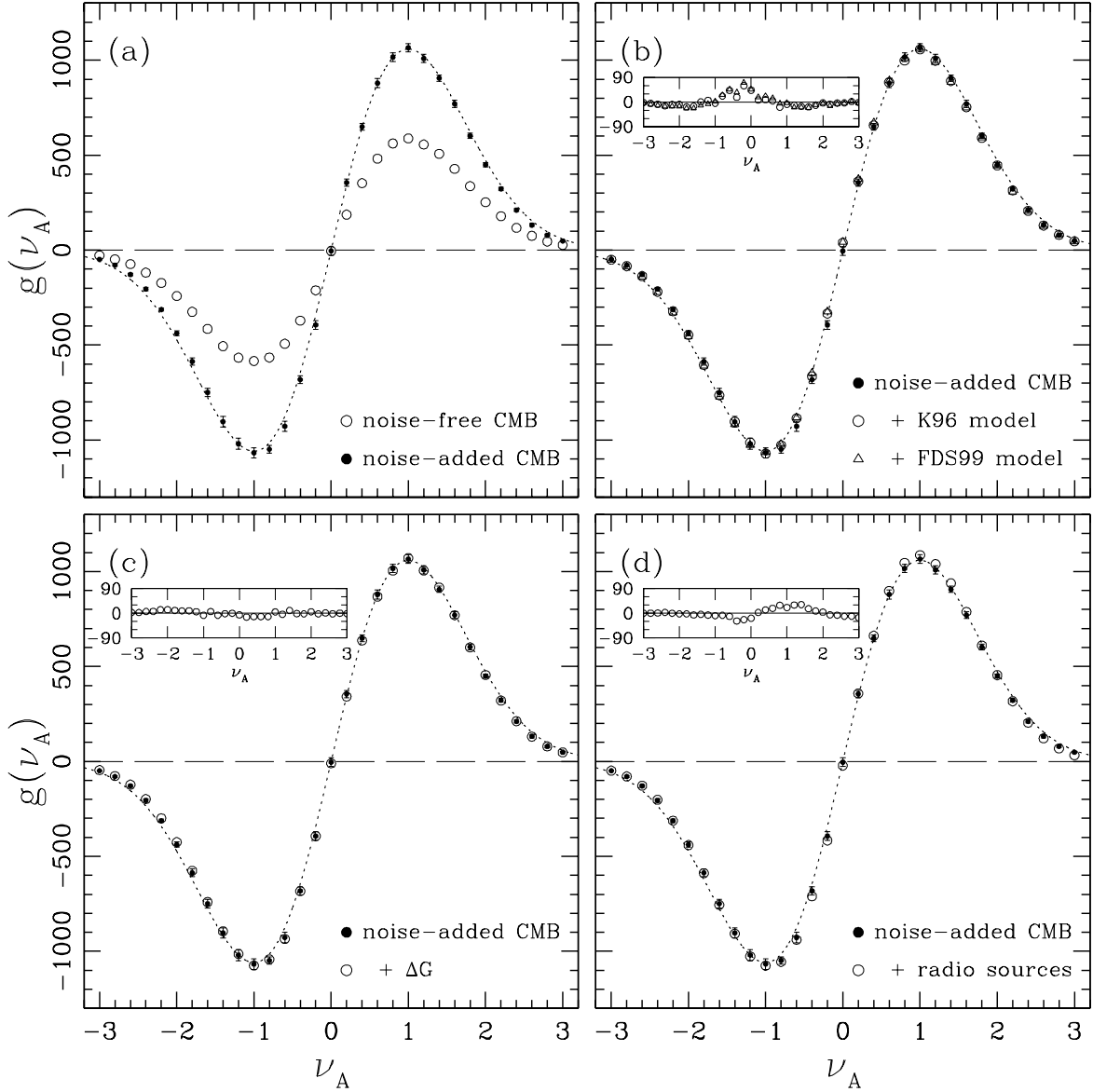}}
\caption{Two-dimensional genus per steradian measured from the noise-added 
         $\Lambda$CDM model CMB anisotropy map ($0\fdg3$ FWHM, $|b|>20\deg$) 
         and from map(s) ($a$) without noise and foreground contamination 
         (open circles), ($b$) with noise and Galactic foreground emission 
         (open circles for the K96 model, open triangles for the FDS99 model), 
         ($c$) with noise and the Galactic emission model uncertainty (open 
         circles), and ($d$) with noise and radio point source emission
         (open circles).  Filled circles in each panel are the genus measured 
         from the noise-added CMB map, with error bars derived from the 
         variance of the genus computed independently in each of the eight 
         octants of the sphere. Dotted curve in each panel shows the functional 
         form expected for a random-phase Gaussian field, $N \nu e^{-\nu^2 
         /2}$, and has been fitted to the filled circles by adjusting $N$. 
         Small panels in the upper-left region of the large panels 
         from ($b$) to ($d$) show genus deviations of the foreground-added
         CMB maps relative to the noise-added CMB result.
         Genus of the CMB map contaminated by FIR extragalactic source 
         emission is almost the same as the foreground-free result and is not 
         shown here (see Table 3).}
\end{figure}
\clearpage 


\clearpage


\begin{thebibliography}{}
\bibitem[Becker et al. (1991)]{becker91}Becker, R.H., White, R.L., \& Edwards, 
   A.L. 1991, ApJS, 75, 1
\bibitem[Bennett et al. (1992)]{bennett92}Bennett, C.L., et al. 1992, ApJ, 396, 
   L7 
\bibitem[Bennett et al. (1994)]{bennett94}Bennett, C.L., et al. 1994, ApJ, 436, 
   423
\bibitem[Bennett et al. (1996)]{bennett96}Bennett, C.L., et al. 1996, ApJ, 464, 
   L1
\bibitem[Boughn et al. (1992)]{boughn92}Boughn, S.P., Cheng, E.S., Cottingham, 
   D.A., \& Fixsen, D.J. 1992, ApJ, 391, L49
\bibitem[Brandt et al. (1994)]{brandt94}Brandt, W.N., Lawrence, C.R., Readhead,
   A.C.S., Pakianathan, J.N., \& Fiola, T.M. 1994, ApJ, 424, 1 
\bibitem[Bouchet \& Gispert (1999)]{bouchet99}Bouchet, F.R., \& Gispert, R. 
   1999, NewA, 4, 443
\bibitem[{\it COBE} DIRBE Exp. Supp. (1998)]{cobe98}{\it COBE} DIRBE 
   Explanatory Supplement, Version 2.3, 1998, ed. M.G. Hauser, T. Kelsall, 
   D. Leisawitz, \& J. Weiland ({\it COBE} Ref. Pub. No. 98-A) (Greenbelt: 
   NASA/GSFC)
\bibitem[Coble et al. (1999)]{coble99} Coble, K., et al. 1999, \apj, 519, L5
\bibitem[Coles (1988)]{coles88}Coles, P. 1988, MNRAS, 234, 509
\bibitem[Coles \& Barrow (1987)]{coles87}Coles, P., \& Barrow, J.D. 1987,
   MNRAS, 228, 407
\bibitem[Coles \& Plionis (1991)]{coles91}Coles, P., \& Plionis, M. 1991,
   MNRAS, 250, 75
\bibitem[Colley (1997)]{colley97}Colley, W.N. 1997, ApJ, 489, 471
\bibitem[Colley et al. (1996)]{colley96}Colley, W.N., Gott, J.R., \& Park, C.
   1996, MNRAS, 281, L82
\bibitem[Danese et al. (1987)]{danese87}Danese, L., Franceschini, A., 
   Toffolatti, L., \& de Zotti, G. 1987, ApJ, 318, L15
\bibitem[de Oliveira et al. (2000)]{oliveira00}de Oliveira-Costa, A., et al.
   2000, ApJ, submitted (astro-ph/0010527)
\bibitem[de Oliveira et al. (1997)]{oliveira97}de Oliveira-Costa, A., Kogut, 
   A., Devlin, M.J., Netterfield, C.B., Page, L.A., \& Wollack, E.J. 1997, 
   ApJ, 482, L17
\bibitem[de Oliveira et al. (1998)]{oliveira98}de Oliveira-Costa, A., Tegmark, 
   M., Page, L.A., \& Boughn, S.P. 1998, ApJ, 509, L9
\bibitem[Dodelson \& Stebbins (1994)]{dodelson94}Dodelson, S., \& Stebbins, A.
   1994, ApJ, 433, 440
\bibitem[Douspis et al. (2001)]{douspis01}Douspis, M., Bartlett, J.G.,
   Blanchard, A., \& Le Dour, M. 2001, A\&A, 368, 1 
\bibitem[Draine \& Lazarian (1998)]{draine98}Draine, B.T., \& Lazarian, A. 1998,
   ApJ, 494, L19
\bibitem[Finkbeiner et al. (1999)]{FDS99}Finkbeiner, D.P., Davis, M., \&
   Schlegel, D.J. 1999, ApJ, 524, 867 (FDS99)
\bibitem[Fisher et al. (1995)]{fisher95}Fisher, K.B., Huchra, J.P., Strauss, 
   M.A., Davis, M., Yahil, A., \& Schlegel, D. 1995, ApJS, 100, 69
\bibitem[Fixsen et al. (1996)]{fixsen96}Fixsen, D.J., Cheng, E.S., Gales, J.M.,
   Mather, J.C., Shafer, R.A., \& Wright, E.L. 1996, ApJ, 473, 576 
\bibitem[Ganga et al. (1997)]{ganga97} Ganga, K., Ratra, B., Gundersen, J.O.,
   \&\ Sugiyama, N. 1997, \apj, 484, 7
\bibitem[Ganga et al. (1998)]{ganga98} Ganga, K., Ratra, B., Lim, M.A., 
   Sugiyama, N., \&\ Tanaka, S.T. 1998, ApJS, 114, 165
\bibitem[Gaustad et al. (1996)]{gaustad96}Gaustad, J.E., McCullough, P.R., \&
   van Buren, D. 1996, PASP, 108, 351
\bibitem[Gawiser et al. (1998a)]{gawiser98a}Gawiser, E., Jaffe, A., \& Silk, J. 
   1998a, AAS, 192, 1703
\bibitem[Gawiser et al. (1998b)]{gawiser98b}Gawiser, E., et al. 1998b, 
   astro-ph/9812237
\bibitem[Gawiser \& Silk (2000)]{gawiser00}Gawiser, E., \& Silk, J. 2000,
   Phys. Rep., 333, 245
\bibitem[Gawiser \& Smoot (1997)]{gawiser97}Gawiser, E., \& Smoot, G.F. 1997, 
   ApJ, 480, L1
\bibitem[G\'orski et al. (1996)]{gorski96}G\'orski, K.M., Banday, A.J., 
   Bennett, C.L., Hinshaw, G., Kogut, A., Smoot, G.F., \& Wright, E.L. 1996,
   ApJ, 464, L11
\bibitem[G\'orski et al. (1998)]{gorski98} G\'orski, K.M., Ratra, B., Stompor,
   R., Sugiyama, N., \& Banday, A.J. 1998, ApJS, 114, 1
\bibitem[Gott et al. (1992)]{gott92}Gott, J.R., Mao, S., Park, C., \& Lahav, O.
   1992, ApJ, 385, 26
\bibitem[Gott et al. (1990)]{gott90}Gott, J.R., Park, C., Juszkiewicz, R.,
   Bies, W.E., Bennett, D.P., Bouchet, F.R., \& Stebbins, A. 1990, ApJ, 352, 1
\bibitem[Griffith \& Wright (1993)]{griffith93}Griffith, M.R., \&  Wright, A.E.
   1993, AJ, 105, 1666
\bibitem[Gundersen et al. (1995)]{gundersen95}Gundersen, J.O., et al. 1995, 
   \apj, 443, L57
\bibitem[Halverson et al. (2001)]{halverson01}Halverson, N.W., et al. 2001, 
   \apj, submitted (astro-ph/0104489)
\bibitem[Hamilton \& Ganga (2001)]{hamilton01}Hamilton, J.-Ch., \&  Ganga, 
   K.M. 2001, A{\&}A, 368, 760
\bibitem[Haslam et al. (1981)]{haslam81}Haslam, C.G.T., Klein, U., Salter, 
   C.J., Stoffel, H., Wilson, W.E., Cleary, M.N., Cooke, D.J., \& Thomasson, P.
   1981, A\&A, 100, 209
\bibitem[Helou (1986)]{helou86}Helou, G. 1986, ApJ, 311, L33
\bibitem[Jaffe et al. (1999)]{jaffe99}Jaffe, A.H., et al. 1999, in ASP Conf.
   Ser. 181, Microwave Foregrounds, ed. A. de Oliveira-Costa \& M. Tegmark
   (San Francisco: ASP), 367
\bibitem[Kamionkowski \& Kosowsky (1999)]{kamionkowski99} Kamionkowski, M.,
   \& Kosowsky, A. 1999, Ann. Rev. Nucl. Part. Sci., 49, 77
\bibitem[Knox \& Page (2000)]{knox00} Knox, L., \& Page, L. 2000, Phys. Rev.
   Lett., 85, 1366
\bibitem[Kogut (1999)]{kogut99}Kogut, A. 1999, in ASP Conf. Ser. 181, Microwave
   Foregrounds, ed. A. de Oliveira-Costa \& M. Tegmark (San Francisco: ASP), 91
\bibitem[Kogut et al. 1996a]{kogut96a}Kogut, A., Banday, A.J., Bennett, C.L., 
   G\'orski, K.M., Hinshaw, G., \& Reach, W.T. 1996a, ApJ, 460, 1
\bibitem[Kogut et al. 1996b]{K96}Kogut, A., Banday, A.J., Bennett, C.L., 
   G\'orski, K.M., Hinshaw, G., Smoot, G.F., \& Wright, E.L. 1996b, ApJ, 
   464, L5 (K96) 
\bibitem[Kogut et al. 1996c]{kogut96c}Kogut, A., Banday, A.J., Bennett, C.L., 
   G\'orski, K.M., Hinshaw, G., Smoot, G.F., \& Wright, E.L. 1996c, ApJ, 
   464, L29
\bibitem[K\"uh et al. (1981)]{kuhr81}K\"uhr, H., Witzel, A., Pauliny-Toth, 
   I.I.K., \& Nauber, U. 1981, A\&AS, 45, 367
\bibitem[Lee et al. (2001)]{lee01}Lee, A.T., et al. 2001, \apj, 561, L1
\bibitem[Leitch et al. (2000)]{leitch00} Leitch, E.M., Readhead, A.C.S., 
  Pearson, T.J., Myers, S.T., Gulkis, S., \& Lawrence, C.R. 2000, 
  \apj, 532, 37
\bibitem[Lim et al. 1996]{lim96}Lim, M.A., et al. 1996, \apj, 469, L69
\bibitem[Melott et al. (1989)]{melott89}Melott, A.L., Cohen, A.P., Hamilton,
   A.J.S., Gott, J.R., \& Weinberg, D.H. 1989, ApJ, 345, 618
\bibitem[Mukherjee et al. (2001)]{mukherjee01}Mukherjee, P., Jones, A.W., 
   Kneissl, R., \& Lasenby, A.N. 2001, MNRAS, 320, 224
\bibitem[Netterfield et al. (2001)]{netterfield01}Netterfield, C.B., et al.
   2001, \apj, submitted (astro-ph/0104460)
\bibitem[Netterfield et al. (1997)]{netterfield97}Netterfield, C.B., Devlin, 
   M.J., Jarosik, N., Page, L., \& Wollack, E.J. 1997, ApJ, 474, 47
\bibitem[Odenwald et al. (1998)]{odenwald98}Odenwald, S., Newmark, J., \& 
   Smoot, G. 1998, ApJ, 500, 554
\bibitem[Park et al. (1998)]{P98}Park, C., Colley, W.N., Gott, J.R., Ratra, 
   B., Spergel, D.N., \& Sugiyama, N. 1998, ApJ, 506, 473 (P98)
\bibitem[Park et al. (2001a)]{park01a}Park, C., Gott, J.R., \& Choi, Y.J. 2001a,
   ApJ, 553, 33
\bibitem[Park et al. (1992)]{park92}Park, C., Gott, J.R., Melott, A.L., \&
   Karachentsev, I.D. 1992, ApJ, 387, 1
\bibitem[Park et al. (2001b)]{park01b}Park, C.-G., Park, C., Ratra, B., \& 
   Tegmark, M. 2001b, ApJ, 556, 582
\bibitem[Platania et al. (1998)]{platania98}Platania, P., Bensadoun, M., 
   Bersanelli, M., de Amici, G., Kogut, A., Levin, S., Maino, D., \& Smoot, 
   G.F. 1998, ApJ, 505, 473
\bibitem[Podariu et al. (2001)]{podariu01} Podariu, S., Souradeep, T., Gott, 
   J.R., Ratra, B., \& Vogeley, M.S. 2001, ApJ, 559, 9
\bibitem[Pryke et al. (2001)]{pryke01}Pryke, C., Halverson, N.W., Leitch, E.M., 
   Kovac, J., Carlstrom, J.E., Holzapfel, W.L., \& Dragovan, M. 2001, \apj,
   submitted (astro-ph/0104490)
\bibitem[Ratra et al. (1999a)]{ratra99a} Ratra, B., Ganga, K., Stompor, R., 
   Sugiyama, N., de Bernardis, P., \&\ G\'orski, K.M. 1999a, ApJ, 510, 11
\bibitem[Ratra et al. (1999b)]{ratra99b} Ratra, B., Stompor, R., Ganga, K., 
   Rocha, G., Sugiyama, N., \&\ G\'orski, K.M. 1999b, ApJ, 517, 549 
\bibitem[Ratra et al. (1997)]{ratra97} Ratra, B., Sugiyama, N., Banday, A.J., 
   \& G\'orski, K.M. 1997, \apj, 481, 22
\bibitem[Reach et al. (1995)]{reach95}Reach, W.T., et al. 1995, ApJ, 451, 188
\bibitem[Refregier et al. (2000)]{refregier00}Refregier, A., Spergel, D.N., 
   \& Herbig, T. 2000, ApJ, 531, 31
\bibitem[Reich \& Reich (1988)]{reich88}Reich, P., \& Reich, W. 1988, A\&AS, 
   74, 7
\bibitem[Rocha (1999)]{rocha99b} Rocha, G. 1999, in Dark Matter in Astrophysics 
  and Particle Physics 1998, ed. H.V. Klapdor-Kleingrothaus \& L. Baudis 
  (Bristol: Institute of Physics Publishing), 238
\bibitem[Rocha et al. (1999)]{rocha99a} Rocha, G., Stompor, R., Ganga, K., 
  Ratra, B., Platt, S.R., Sugiyama, N., \& G\'orski, K.M. 1999, ApJ, 525, 1
\bibitem[Schlegel et al. (1998)]{schlegel98}Schlegel, D.J., Finkbeiner, D.P., 
   \& Davis, M. 1998, ApJ, 500, 525
\bibitem[Simonetti et al. (1996)]{simonetti96}Simonetti, J.H., Dennison, B., 
   \& Topasna, G.A. 1996, ApJ, 458, L1
\bibitem[Smith et al. (1987)]{smith87}Smith, B.J., Kleinmann, S.G., Huchra, 
   J.P., \& Low, F.J. 1987, ApJ, 318, 161
\bibitem[Smoot et al. (1994)]{smoot94}Smoot, G.F., Tenorio, L., Banday, A.J.,
   Kogut, A., Wright, E.L., Hinshaw, G., \& Bennett, C.L. 1994, ApJ, 437, 1
\bibitem[Soifer et al. (1989)]{soifer98}Soifer, B.T., Boehmer, L., Neugebauer, 
   G., \& Sanders, D.B. 1989, AJ, 98, 766
\bibitem[Sokasian et al. (1998)]{sokasian98}Sokasian, A., Gawiser, E., \&  
   Smoot, G.F. 1998, astro-ph/9811311
\bibitem[Stompor et al. (2001)]{stompor01}Stompor, R., et al. 2001, \apj, 
   561, L7
\bibitem[Strauss et al. (1990)]{strauss90}Strauss, M.A., Davis, M., Yahil, A., 
   \& Huchra, J.P. 1990, ApJ, 361, 49
\bibitem[Tegmark \& Efstathiou (1996)]{tegmark96}Tegmark, M., \& Efstathiou, G.
   1996, MNRAS, 281, 1297
\bibitem[Tegmark et al. (2000)]{tegmark00}Tegmark, M., Eisenstein, D.J., Hu, W.,
   \& de Oliveira-Costa, A. 2000, ApJ, 530, 133
\bibitem[Toffolatti et al. (1998)]{toffolatti98}Toffolatti, L., Arg\"ueso 
   G\'omez, F., De Zotti, G., Mazzei, P., Franceschini, A., Danese, L., \& 
   Burigana, C. 1998, MNRAS, 297, 117
\bibitem[Wang \& Mathews (2000)]{wang00}Wang, Y., \& Mathews, G. 2000, \apj,
   submitted (astro-ph/0011351)
\bibitem[White \& Becker (1992)]{white92}White, R.L., \& Becker, R.H. 1992, 
   ApJS, 79, 331
\bibitem[Wright et al. (1991)]{wright91}Wright, E.L., et al. 1991, ApJ, 381, 200
\bibitem[Xu et al. (2001)]{xu01}Xu, Y., Tegmark, M., \& de Oliveira-Costa, A.
   2001, Phys. Rev. D, submitted (astro-ph/0104419)
\end{thebibliography}
\end{document}